\documentclass[12pt,tightenlines,nofootinbib,superscriptaddress,notitlepage, APS, pra]{revtex4-1}

\usepackage{amsfonts,amssymb,amsthm}
\usepackage{appendix}

\usepackage{amssymb,amstext,amsmath,amsthm}
\usepackage{amsfonts}
\usepackage{color}
\usepackage{hyperref}
\usepackage{float}
\usepackage{placeins}
\usepackage{bbold}
\usepackage{indentfirst}
\usepackage{tabu}
\usepackage{verbatim}
\usepackage{comment}

\newcommand{\be}{\begin{equation}}
\newcommand{\ee}{\end{equation}}
\newcommand{\barray}{\begin{array}}
\newcommand{\earray}{\end{array}}
\newcommand{\bea}{\begin{eqnarray}}
\newcommand{\eea}{\end{eqnarray}}
\newcommand{\bs}{\begin{subequations}}
\newcommand{\es}{\end{subequations}}
\newcommand{\balign}{\begin{align}}
\newcommand{\ealign}{\end{align}}
\newcommand{\equ}{\begin{equation}}
\newcommand{\nequ}{\end{equation}}
\newcommand{\eqa}{\begin{eqnarray}}
\newcommand{\neqa}{\end{eqnarray}}

\def\sig{\sigma}

\def\eps{\epsilon}

\def\lam{\lambda}

\def\eps{\epsilon}

\def\pa{\phantom{\alpha}}
\def\bs{\bar{\sigma}}

\newcommand{\rd}{\mathrm{d}}

\newcommand{\N}{\nabla}

\def\sl2c{SL(2,\mathbb{C})}

\newcommand{\mc}[1]{\mathcal{#1}}

\def\bs{\bar{s}}

\usepackage{bbm}

\usepackage{tikz}
\usetikzlibrary{decorations.pathmorphing}
\usepackage{tkz-euclide}
\usetikzlibrary{decorations.markings}
\tikzset{->-/.style={decoration={
  markings,
  mark=at position .5 with {\arrow{>}}},postaction={decorate}}}

\begin{document}
\title{\Large{Action and Vertices in the Worldine Formalism}}

\author{{Laurent Freidel}}
\smallskip 
\affiliation{{Perimeter Institute for Theoretical Physics, Waterloo, Ontario, Canada}}
\author{{Trevor Rempel}}

\affiliation{{Perimeter Institute for Theoretical Physics, Waterloo, Ontario, Canada}}
\affiliation{\small\textit{Department of Physics, University of Waterloo, Waterloo, Ontario, Canada}}
\smallskip
\date{ \today}
\bigskip
\begin{abstract}
Utilizing the worldline formalism we study the effects of demanding local interactions on the corresponding vertex factor. We begin by reviewing the familiar case of a relativistic particle in Minkowksi space, showing that localization gives rise to the standard conservation of momentum at each vertex. A generalization to curved geometry is then studied and a notion of covariant Fourier transform is introduced to aid in the analysis. The vertex factor is found to coincide with the one derived for flat spacetime. Next, we apply this formalism to a loop immersed in a gravitational field, demonstrating that the loop momenta is determined entirely by the external momenta. Finally, we postulate that the semi--classical effects of quantum gravity on the Feynman path integral can be accounted for by a modification to the vertex factor which de--localizes the vertex. We study one particular Lorentz invariant de--localization which, remarkably, has no effect on conservation of vertex momenta.
\end{abstract}
\maketitle

\section{Introduction}
The principle of locality states that interactions occur only if there is a coincidence of multiple particles at a single point in spacetime. This is an intuitive notion and has been taken for granted since Maxwell's introduction of fields to describe electromagnetism and Einstein's prohibition against superluminal exchange of information. 
The assumption of locality is not inviolable though, in fact, relative locality (see \cite{relativelocality},\cite{gammarayburst},\cite{deepening}) proposes a systematic weakening of absolute locality by allowing momentum space to posses a non--trivial geometry. The theory predicts that localization is an observer dependent phenomena with the degree of localization scaling with distance from the interaction point. Furthermore, the authors have shown, see \cite{scalarQFT}, that this non--locality can be entirely absorbed into a de--localisation of the  interaction term in scalar QFT. A key question then arises, can one link this vertex de--localization to quantum gravity effects? The current paper takes a preliminary step towards answering this question by studying the coupling of a particle worldline to a gravitational field and investigating its key properties. More specifically, we utilize the worldline formalism (see \cite{worldlineoverview} for a review of the formalism along with a list of references) to study the behaviour of scalar particles undergoing an arbitrary series of interactions while propagating through an arbitrary spacetime geometry. We are aware of a similar investigation in the context of relative locality by  J. Kowalski--Glikman et al. \cite{jerzy}.   \\
\indent The paper is organized as follows. We begin by considering the standard example of the relativistic particle in a flat spacetime since this allows us to introduce the relevant techniques and make a smooth transition to a more general geometry. Having written down the action for a relativistic particle propagating along its worldline we observe that the amplitude for propagation from one point in spacetime to another is the path integral of the exponential of this action. An arbitrary scattering process is then considered, we place no restrictions on the form of the interactions other than demanding locality. Following a brief calculation we find the expected result, edge momenta is constant, momentum is conserved at each vertex and edge momenta is identical to vertex momenta.\\
\indent After exhausting this familiar example we generalize the worldline action to allow for non-trivial spacetime geometries. We begin by demonstrating that edge momenta are no longer constant, instead they are carried along the worldline by parallel transport. Next, we introduce a notion of covariant Fourier transform which is then utilized in deriving the vertex factor. We find no modification from the case of flat spacetime, momentum is conserved at each vertex and edge momenta is identical to vertex momenta.\\
\indent These results are then used to analyse a loop diagram in the presence of a gravitational field. We show that in a generic curved spacetime the loop momenta is entirely determined by the external momenta, presenting an intriguing approach for regulating the ultraviolet divergences which plague loop integrals in standard QFT. Finally, we argue that the semi--classical effects of quantum gravity can be accounted for by modifying the interaction vertex so as to relax strict locality. We then make a particular choice for the de--localized vertex which preserves Lorentz invariance and demonstrate, rather remarkably, that conservation of momentum at a vertex is preserved. 
\section{Worldline Action in Minkowksi Space}
Consider a particle of mass $m$ propagating in a spacetime with flat Minkowksi metric $\eta_{\mu\nu}$ and having a worldline given by $X^\mu(\tau)$, for some parameter $\tau$. The motion of such a particle is governed by the action
\begin{equation}\label{gaction}
S[e,X] = \frac{1}{2}\int d\tau \left(\frac{1}{e}\dot{X}^\mu\dot{X}^\nu \eta_{\mu\nu} - e m^2\right),
\end{equation}
where $\dot{X}^\mu = dX^\mu/d\tau$ and $-e^2(\tau)$ is the metric along the worldline. Under a change in parametrization $\tau \to s(\tau)$ we have $e(\tau) \to \tilde{e}(s) = (d\tau/ds) e(\tau)$ and so the worldline metric ensures that $S[e,X]$ is invariant under such re-parametrizations. \\
\indent It will prove convenient to re--write this action in--terms of the momentum conjugate to $X^\mu(\tau)$, which we easily calculate to be
\begin{align*}
 P_\mu = \frac{\partial \mathcal{L}}{\partial \dot{X}^\mu} = \frac{1}{e}\dot{X}^\nu \eta_{\mu\nu}.
\end{align*}
On the other hand, taking the variation of $S[e,X]$ with respect to $e$ gives the constraint $X^2/e^2+m^2 = 0$, and upon substituting for $P_\mu$ we obtain the standard mass-shell condition
\begin{align}\label{motion}
P^2 +m^2 = 0.
\end{align}
\indent A brief calculation shows that the Hamiltonian for this system is $\mathcal{H} = e(P^2 +m^2)/2$; an inverse Legendre transform then gives $\mathcal{L} = \dot{X}^\mu P_\mu - \mathcal{H}$ as the Lagrangian. Noting that the action \eqref{gaction} is just the time integral of the Lagrangian we find
\begin{align}\label{actionXP}
S[X,P,e] = \int_0^1 d\tau \left(\dot{X}^\mu P_\mu - \frac{e}2(P^2 + m^2)\right).
\end{align}
In this formulation the worldline metric behaves like a Lagrange multiplier that imposes the mass shell constraint \eqref{motion}. It is conventional to re--label the worldline metric as the lapse function, $e(\tau) = N(\tau)$, so that \eqref{actionXP} becomes
\begin{align}\label{final}
S[X,P,N] &=\int_0^1 d\tau \left[\dot{X}^\mu(\tau) P_\mu(\tau) - \frac{N(\tau)}{2}\left(P^2(\tau)+m^2\right)\right] .
\end{align}
\indent Suppose that the worldline of the particle satisfies $X^\mu(0) = y^\mu$ and $X^\mu(1) = x^\mu$, i.e. the worldline begins at the point $y$ and terminates at the point $x$.  The amplitude for propagating from $y$ to $x$ is then obtained by taking the path integral of the exponential of the action \eqref{final}, viz
\begin{equation}\label{G}
\begin{aligned}
G(x,y) &= \int\mathcal{D}X\mathcal{D}P\mathcal{D}N\exp\left\lbrace i\int_0^1 d\tau \left[\dot{X}^\mu(\tau) P_\mu(\tau) - \frac{N(\tau)}2\left(P^2(\tau)+m^2\right)\right] \right\rbrace.
\end{aligned}
\end{equation}
$G(x,y)$ is simply the propagator for the theory and so we will represent it graphically in the usual way
\begin{equation*}
G(x,y) = 
\begin{tikzpicture}
[place/.style={circle, draw= black, fill=black, inner sep =2pt}]
\node[place] (a) at (0,0) [label = above: $x$] {};
\node[place] (b) at (2,0)[label = above: $y$]   {};
\draw[->, thick] (a) to node {} node {} (b);
\end{tikzpicture},
\end{equation*}
where the arrow indicates the direction of momentum flow.

\section{Propagation Amplitude for an Arbitrary Process}\label{sec: FlatSpace}
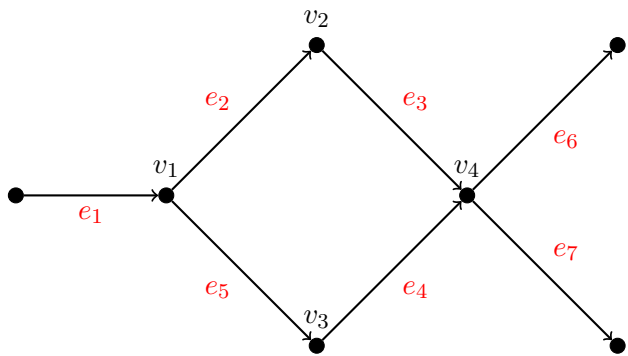
\begin{figure}[H]
\begin{center}
\begin{tikzpicture}
[place/.style={circle, draw= black, fill=black, inner sep =2pt}]
\node[place] (a) at (0,0) {};
\node[place] (b) at (2,0)[label = above: $v_1$]   {};
\node[place] (c) at (4,2)[label = above: $v_2$]  {};
\node[place] (d) at (4,-2)[label = above: $v_3$]  {};
\node[place] (e) at (6,0) [label = above: $v_4$] {};
\node[place] (f) at (8,2) {};
\node[place] (g) at (8,-2) {};
\draw[->, thick] (a) to node {} node [auto, swap, red] {$e_1$} (b);
\draw[<-, thick] (c) to node {} node [auto, swap, red] {$e_2$} (b);
\draw[<-, thick] (e) to node {} node [auto, swap, red] {$e_3$} (c);
\draw[->, thick] (d) to node {} node [auto, swap, red] {$e_4$} (e);
\draw[->, thick] (b) to node {} node [auto, swap, red] {$e_5$} (d);
\draw[->, thick] (e) to node {} node [auto, swap, red] {$e_6$} (f);
\draw[<-, thick] (g) to node {} node [auto, swap, red] {$e_7$} (e);
\end{tikzpicture}
\caption{\textit{A possible graph, $\Gamma$, with $n_i = 1$ and $n_f = 2$.}}
\label{fig:graph}
\end{center}
\end{figure}

Consider a process in which $n_i$ initial state particles undergo a series of interactions to produce $n_f$ final state particles. No restriction is placed on the number of particles participating in a given interaction, we demand only locality, i.e. interacting particles occupy a single point in spacetime. This evolution can be represented by an oriented graph $\Gamma$ in which the edges, labelled by $e$, represent the worldlines of the particles and the vertices, labelled by $v$, represent interactions. An example with $n_i =1$ and $n_f=2$ is given in Figure \ref{fig:graph}.\footnote{To emphasize, a vertex is a point having both incoming \emph{and} outgoing momentum. Therefore, where an initial edge originates and where a final edge terminates are not considered vertices.}

\indent Let $X_e(\tau)$ denote the worldline of a particle propagating along the edge $e$ and $P_e(\tau)$ the momentum it carries. If we reverse the orientation of an edge, $e \to -e$, then $X_{e}(\tau) = X_{-e}(1-\tau)$ since  $X_{-e}(\tau)$ traverses the same path as $X_e(\tau)$ only backwards. Similarly, $P_e (\tau) = -P_{-e}(1-\tau)$, where the minus sign takes into account that the direction of momentum flow has be reversed. We will adopt the notation $x_e \equiv X_e(0)$ and $x_{-e} \equiv X_{-e}(0)= X_e(1)$ for the  endpoints of the edge $e$ while $x_e^\mathrm{in}$ and $x_{-e}^\mathrm{out}$ will denote the coordinates of the initial and final state particles respectively. The amplitude for the graph $\Gamma$, denoted $I_\Gamma(x_e^\mathrm{in}, x_{-e}^\mathrm{out})$, is constructed as follows: 1) Introduce vertex coordinates $z_v$ 2) Assign a propagator to each edge $e$ and then form their product 3) Integrate over the fiducial coordinates $z_v$. Implementing this procedure yields
\begin{equation}\label{Idef}
I_\Gamma(x_e^\mathrm{in}, x_{-e}^\mathrm{out}) = \int\prod_{v \in \Gamma} d^4z_v\prod_{\mathrm{initial}\; e} G_e\left(x_e, z_{e,t}\right) \prod_{\mathrm{internal}\; e}G_e\left(z_{e,s}, z_{e,t}\right) \prod_{\mathrm{final}\; e}G_e\left(z_{e,s}, x_{-e}\right),
\end{equation}
where $z_{e,s}$ and $z_{e,t}$ are, respectively, the coordinates of the sourcing and terminating vertex of the edge $e$. The requirement that interactions occur at a single point in spacetime can be made explicit by extracting a delta function for each vertex and re--writing \eqref{Idef} as
\begin{align}\label{iglocal}
I_\Gamma &=\int \prod_{v\in \Gamma} d^4z_v \prod_{v\in \Gamma}\prod_{\substack{{s_e = v}\\{t_e = v}}} d^4 x_e \delta^{(4)}\left(x_e - z_v\right)\prod_{e\in \Gamma} G_e\left(x_e, x_{-e}\right),
\end{align}
where the product $\prod_{\substack{{s_{e}=v}\\{t_e = v}}}$ is taken over all edges sourcing ($s_e$) from $v$ and terminating ($t_e$) at $v$ with the latter having their orientation reversed. For example, referring to Figure \ref{fig:graph} we have
\begin{equation*}
\prod_{\substack{{s_e = v_4}\\{t_e = v_4}}} d^4x_e = d^4x_{-e_3}d^4x_{-e_4}d^4x_{e_6}d^4x_{e_7}.
\end{equation*}
This type of product will appear repeatedly and it will be convenient to introduce the notation
\begin{align*}
\prod_{v\in \Gamma}\prod_{\substack{{s_e = v}\\{t_e = v}}}\equiv \prod_{v,e}.
\end{align*}
\indent Returning to our expression for $I_\Gamma$ in equation \eqref{iglocal} we take the Fourier transform of the delta functions and expand the $G_e$ using equation \eqref{G} from the previous section. The result is
\begin{align}
I_\Gamma &= \int\prod_{v\in \Gamma}  d^4z_v \prod_{v,e} d^4 x_e\frac{d^4p_e}{(2\pi)^4}\prod_{e\in \Gamma}\mathcal{D}\mu_e \exp\left(-i S_\Gamma\right)\label{igamma},
\end{align}
where $\mathcal{D}\mu_e = \mathcal{D}X_e\mathcal{D}P_e\mathcal{D}N_e$ and
\begin{equation}
\begin{aligned}\label{action}
S_\Gamma &= -\sum_{e\in \Gamma}\int_0^1 d\tau \left[\dot{X}_e \cdot{P}_{e} - N_e\left(P_e^2 + m_e^2\right)\right]  + \sum_{v,e} p_e\cdot (x_e - z_v).
\end{aligned}
\end{equation}
The coordinates, $p_e$, employed in the Fourier transform are dual to the vertex coordinates $z_v$, a relationship which suggests the designation ``vertex momentum'' for the $p_e$. This should be contrasted with the $P_e(\tau)$ which are dual to the worldline coordinates $X_e(\tau)$ and referred to as edge momenta. \\
\indent To obtain the equations of motion for this system, and in particular the vertex factor, we simply take the variation of $S_\Gamma$:
\begin{align*}
\delta S_\Gamma &= \sum_{e\in \Gamma}\int_0^1 d\tau \left[\delta{X}^\mu_e\dot{P}_{\mu,e} - \dot{X}^\mu_e  \delta P_{\mu,e} -\delta N_e \left(P_e^2 + m_e^2\right) + N_e P^\mu_{e}\delta P_{\mu,e}\right]  \\
&\qquad + \sum_{v,e} \left[(\delta p_{\mu,e})(x^\mu_e - z^\mu_v) + p_{\mu,e}\delta x^\mu_e - p_{\mu,e}\delta z^\mu_v - P_{\mu,e}(0)\delta x_e^\mu\right],
\end{align*} 
where we have assumed that $\delta X_e(0) = 0$ and $\delta X_e(1)=0$ for incoming and outgoing edges respectively. Setting the variations along the worldline to zero we obtain
\begin{gather}
\dot{P}_{\mu,e} = 0 \qquad \dot{X}^\mu_e = N_e\eta^{\mu\nu}P_{\nu,e} \qquad P_e^2 + m_e^2 = 0, \label{flatkin}
\end{gather}
which hold for all $e \in \Gamma$. The interpretation is standard; momentum is conserved along a linear worldline and the mass--shell condition is satisfied. Turning now to the variations at the vertices we find
\begin{gather}{}
x^\mu_e = z^\mu_v \qquad  \forall v \in \Gamma\label{flatloc},\\
p_{\mu,e} = P_{\mu,e}(0)\qquad   \forall e \in \Gamma\label{flatmom},\\
\sum_{\substack{{s_e = v}\\{t_e = v}}} p_{\mu,e}=0 \qquad \forall v \in \Gamma\label{flatcon},
\end{gather}
and it is assumed that if $x_e$ and $z_v$ appear in the same equation the edge $e$ innervates the vertex $v$. Equation \eqref{flatloc} can be easily recognized as the locality condition; all interactions must occur at a single point in spacetime.  The subsequent equation relates the vertex momenta to the edge momenta, and noting that the edge momenta is conserved we obtain
\begin{align*}
P_e(0) - P_e(1) = p_e + p_{-e} = 0,
\end{align*} 
where $P_e(1) = -P_{-e}(0)$ was used in the second equality. The locality condition can be combined with this relation and the expression for $\dot{X}_e$ in \eqref{flatkin} to relate the vertex momenta to a difference in position, viz
\be
z_{t_{e}}-z_{s_{e}} = \tau_{e} p_{e},\qquad \tau_{e}\equiv \int_{0}^{1}N_{e}(\tau) \rd \tau.
\ee
The interpretation of the final equation, \eqref{flatcon}, is immediate when combined with equation \eqref{flatmom}, we find
\be
\sum_{\substack{{s_e = v}\\{t_e = v}}} P_{\mu,e}(0)=0,
\ee
which expresses the conservation of edge momentum at each vertex. Having exhausted this simplest example we now consider the case where the geometry of spacetime is non-trivial.

\section{Worldline Action in Curved Spacetime}\label{sec:Worldline Curved}
Recall the form of the worldline action for a particle propagating in flat spacetime
\begin{align*}
S[X,P,N] = \int d\tau \left[\dot{X}^a(\tau) P_a(\tau) - \frac{N(\tau)}{2}\left(P_a(\tau)P_b(\tau)\eta^{ab} + m^2\right)\right]
\end{align*}
where the Latin indices $a,b$ indicate that $X^a$ and $P_a$ take values in Minkowski space. We now suppose that $X^a \mapsto X^\mu$ takes values in some generic manifold $\mathcal{M}$ with metric $g_{\mu\nu} = e^a_\mu e^b_\nu \eta^{ab}$. The momentum conjugate to $X^\mu$, say $P_\mu$, takes values in $T_{X(\tau)}\mathcal{M}$, but for convenience we write it in--terms of the flat momentum $P_a$ as $P_\mu= e^a_\mu (X) P_a$. Making the additional replacement $\eta^{ab} \to g^{\mu\nu}$ in the mass shell term we obtain the action
\begin{align}\label{curvedaction}
S[X,P,N] = \int d\tau \left[\dot{X}^\mu e^a_\mu P_a - \frac{N}{2} \left(P_aP_b \eta^{ab} + m^2\right)\right],
\end{align}
where we have used $g^{\mu\nu}e^a_\mu e^b_\nu = \eta^{ab}$. To demonstrate that this action is reasonable we will now calculate the equations of motion for $X^\mu$ and $P_a$:
\begin{align*}
\delta S &= \int d\tau \Big[-\frac{d}{d\tau}\left(e^a_\mu P_a\right)\delta{X}^\mu +\dot{X}^\mu e^a_{\mu,\nu} P_a \delta X^\nu + \dot{X}^\mu e^a_\mu \delta P_a  - \frac{1}{2}\delta N\left(P_aP_b \eta^{ab} + m^2\right) \\
&\qquad  -NP_b \eta^{ab} \delta P_a\Big].
\end{align*}
Setting the variations to zero we find
\begin{gather}
-\frac{d}{d\tau}\left(e^a_\mu P_a\right)  + e^a_{\nu,\mu}P_a \dot{X}^\nu = 0\label{curvedP1},\\
\dot{X}^\mu e^a_\mu - NP_b\eta^{ab} = 0\label{curvedP2},\\
P_aP_b\eta^{ab} + m^2 = 0\label{curvedP3}.
\end{gather}
The second equation can be solved for $P_a$, and after changing variables to proper time $ds = Nd\tau$ we obtain
\begin{align}
P_a = \eta_{ab}e^b_\mu \partial_s{X}^\mu .
\end{align}
Substituting this relation into equation \eqref{curvedP1} gives the evolution equation for $X^\mu$:
\begin{align}\label{Xequation}
\partial_s\left(e^a_\mu e^b_\nu \eta_{ab} \partial_s{X}^\nu\right) - \eta_{ab}e^a_{\nu,\mu} e^b_\alpha\partial_s{X}^\nu \partial_s{X}^\alpha = 0.
\end{align}
The product of tetrads in the second term can be symmetrized over $(\alpha, \nu)$ and re--written as,  $\partial_\mu(\eta_{ab}e^a_\nu e^b_\alpha)\dot{X}^\nu \dot{X}^\alpha/2$. Making the replacement $g_{\mu\nu} = \eta_{ab}e^a_\mu e^b_\nu$ in \eqref{Xequation} then gives
\begin{align*}
\partial^2_s{X}^\rho + \frac{1}{2}g^{\rho\mu}\left(g_{\mu\nu,\alpha} + g_{\mu\alpha,\nu} - g_{\nu\alpha,\mu}\right)\partial_s{X}^\nu\partial_s{X}^\alpha = 0,
\end{align*}
which is just the geodesic equation. This is what we expected, free particles in a curved spacetime obey the geodesic equation. It is also enlightening to write the evolution equation \eqref{curvedP1} in terms of $P_{a}$ as 
\begin{align}
\frac{d}{d s}P_{a} &= e^{\mu}_{a}(e_{\nu,\mu}^{b}-e_{\mu,\nu}^{b})P_{b}\partial_s{X}^{\nu}. 
\end{align}
Noting that the above equation is antisymmetric in $\mu,\nu$, we obtain that $P^{a}P_{a}$ is conserved along the worldline and so the mass--shell constraint is satisfied if it is satisfied initially. Introducing the spin connection, we can write this in the more compact form
\be
\frac{d}{d s}P_{a} - \partial_s{X}^{\mu} \omega_{\mu}{}^{b}{}_{a}P_{b}  = 0, \qquad  \omega_{\mu}{}^{b}{}_{a} \equiv -(\N_{\mu} e^{b}_{\nu}) e^{\nu}_{a}.
\ee
Equations of this type can be solved by iterative integration, viz
\be\label{Pedge}
P_{a}(s) =  P_{b}(0) U(s)^{b}{}_{a},\qquad U(s) =\overrightarrow{\exp}\int_{0}^{s}\rd \tau \dot{X}^{\mu} \omega_{\mu}(X(\tau)),
\ee
where $U(s) $ is the parallel transport operator along the geodesic to which $P_a$ is dual. \\
\indent Returning to our expression for the worldline action in curved spacetime, equation \eqref{curvedaction}, it follows that the amplitude for a particle to propagate from the point $y$ to the point $x$ is given by
\begin{align*}
\mathcal{G}(x,y) = \int \mathcal{D}X\mathcal{D}P\mathcal{D}N \exp\left(i\int_0^1 d\tau \left[\dot{X}^\mu e^a_\mu(X) P_a - N\left(P_aP_b \eta^{ab} + m^2\right)\right]\right).
\end{align*}
The development now proceeds as in the previous section. We consider an arbitrary process in which $n_i$ initial state particles undergo a series of interactions to produce $n_f$ final state particles. No restrictions are placed on these interactions other than demanding locality. The process is represented by an oriented graph $\Gamma$ with a corresponding amplitude given by
\begin{align*}
\mathcal{I}_\Gamma(x_e^\mathrm{in}, x_{-e}^\mathrm{out}) &= \int\prod_{v \in \Gamma} d\mu(z_v) \prod_{\mathrm{initial}\; e} \mc{G}_e\left(x_e, z_{e,t}\right) \prod_{\mathrm{internal}\; e}\mc{G}_e\left(z_{e,s}, z_{e,t}\right) \prod_{\mathrm{final}\; e}\mc{G}_e\left(z_{e,s}, x_{-e}\right),
\end{align*}
where $d\mu(z_v) = \sqrt{g(z_v)}d^4z_v$ is the covariant measure on spacetime. \\
\indent Naively, the form of the interaction vertex can be obtained by proceeding as we did in flat spacetime: Make locality explicit by extracting a delta function for each vertex, Fourier transform the delta function, Take the variation of the resulting action. Unfortunately, the second step in this sequence presents a major impediment. A key property of $\mc{I}_{\Gamma}$ is its invariance under diffeomorphisms, a symmetry which is not preserved by the standard Fourier transform. To proceed we need to define a generalization of the Fourier transform which \textit{does} preserve diffeomorphism covariance; it is to the development of such a ``covariant Fourier transform'' that we turn in the subsequent section.

\section{World Function and Covariant Fourier Transform}
The standard Fourier kernel is given by $\exp(ix\cdot p)$ and manifestly breaks diffeomorphism invariance. In particular, since $x^\mu$ transforms as a coordinate and not a contravariant vector, its contraction with the covariant vector $p_\mu$ does not transform as a scalar. It is, therefore, the failure of $x^\mu$ to behave contravariantly which destroys the covariance of the Fourier kernel. In this section we will present a generalization of $x^\mu$ which does transform properly and then use this ``generalized coordinate'' to construct a covariant Fourier kernel. Our first step is to introduce Synge's world function.

\subsection{Synge's World Function}
Consider a manifold $\mc{M}$ endowed with a metric $g_{\mu\nu}$ and a torsionless metric compatible connection $\Gamma^\mu_{\nu\rho}$. We restrict our attention to an open region $\mathcal{O} \subset \mc{M}$ such that for every $x^\prime,x \in \mathcal{O}$ there is a unique geodesic $\gamma_{x^\prime x}$ connecting $x^\prime$ to $x$. Synge's world function $\sig(x^\prime, x)$ (see \cite{SyngeGR}) is then defined as
\begin{equation}\label{WF}
\sigma(x,x^\prime) \equiv \frac{1}{2}\left(s-s^\prime\right)^2,
\end{equation}
where $s-s^\prime$ is the arc-length between $x^\prime$ and $x$ as determined by $\gamma_{x^\prime x}$. This definition makes clear that the world function is symmetric upon interchange of its arguments and transforms as a scalar with respect to both $x^\prime$ and $x$. We will now derive some important properties of $\sig(x,x')$ which will prove useful in subsequent sections.  \\
\indent Let $\xi^\mu(\lam)$ be an affine parametrization of $\gamma_{x^\prime x}$ so that the geodesic can be described by the Lagrangian
\begin{equation*}
L = \frac{1}{2}g_{\mu\nu}\frac{d \xi^\mu}{d\lambda}\frac{d \xi^\nu}{d\lambda}.
\end{equation*}
Let $\lambda_0$ denote the value of $\lambda$ at $x^\prime$ and $\lambda_1$ its value at $x$, then
\begin{align}\label{arc}
s-s^\prime = \int _{s^\prime}^s ds = \int_{s^\prime}^s \sqrt{g_{\mu\nu}d\xi^\mu d\xi^\nu} = \int_{\lambda_0}^{\lam_1} \sqrt{2L}d\lambda = (\lam_1-\lam_0)\sqrt{2L},
\end{align}
where the final equality follows by noting that $L$ is constant along an affinely parametrized geodesic. Substituting this result into the definition of the world function we obtain 
\begin{align}\label{HPF}
\sigma(x,x^\prime) = L(\lambda_1 - \lambda_0)^2 = (\lambda_1 - \lambda_0)\int_{\lambda_0}^{\lambda_1} L d\lambda =(\lambda_1 - \lambda_0) S(x,\lambda_1 ; x^\prime , \lambda_0),
\end{align}
where $S$ is Hamilton's principle function. The covariant derivatives of $\sig(x,x')$ can now be calculated by means of the Hamilton--Jacobi equation, e.g. taking the covariant derivative at $x$ we find 
\begin{align}\label{sigmu}
\sigma_{;\mu} = \sigma_{,\mu} = (\lambda_1 - \lambda_0)\frac{\partial S}{\partial x^{\mu}} = (\lambda_1 - \lambda_0)\frac{\partial L}{\partial \dot{x}^{\mu}} = (\lambda_1 - \lambda_0)g_{\mu \nu}\dot{x}^{\nu},
\end{align}
where $\dot{x}^\mu = d\xi^\mu/d\lambda |_{\lambda = \lambda_1}$. A short calculation then yields the equation of motion
\begin{equation}\label{EOM1}
\frac{1}{2}g^{\mu\nu}\sigma_{;\mu}\sigma_{;\nu} = \sigma.
\end{equation}
Swapping the roles of $x$ and $x^\prime$ gives similar expressions for the covariant derivative of the worldfunction at $x^\prime$
\begin{gather}
\sigma_{;\mu^\prime} =-(\lambda_1 - \lambda_0) g_{\mu^\prime\nu^\prime}\dot{x}^{\nu^\prime}\label{sigmup}\\
\frac{1}{2}g^{\mu^\prime\nu^\prime}\sigma_{;\mu^\prime}\sigma_{;\nu^\prime} = \sigma \label{EOM2}.
\end{gather}
It should be noted that the indices on a tensor indicate the point at which it is evaluated, for example $g_{\mu'\nu'} = g_{\mu'\nu'}(x')$. We can generate implicit expressions for higher derivatives of the world function by differentiating \eqref{EOM1} and \eqref{EOM2} repeatedly. In particular, taking one additional derivative we find
\begin{align}
\sig^\mu_{\;\; \nu}\sig_\mu = \sig_\nu \qquad \sig^{\mu^\prime}_{\;\;\nu^\prime}\sig_{\mu^\prime} = \sig_{\nu^\prime} \label{SD1}\\
\sig^\mu_{\;\; \nu^\prime}\sig_\mu = \sig_{\nu^\prime} \qquad \sig^{\mu^\prime}_{\;\;\nu} \sig_{\mu^\prime} = \sig_\nu \label{SD2},
\end{align}
where we have omitted the semicolon to simplify notation. Although these equations were easily derived they are quite significant. The first set demonstrates that the second order derivative of the world function at $x$ (or $x'$) behaves like a Kronecker delta when acting on $\sig_\mu$ (or $\sig_{\mu'}$). Referring to equations \eqref{Usig1} and \eqref{Usig2} in Appendix \ref{ap:propagator}, the second set shows that, up to a sign, the second order mixed derivative of $\sig$ behaves like the parallel propagator when acting on $\sig_\mu$ or $\sig_{\nu'}$. \\
\indent Let us now take a brief look at the behaviour of $\sig(x,x')$ when $x$ tends to $x'$, or vice versa. This is known as the coincidence limit and is obtained by letting $\lam_1 \to \lam_0$ on either side of \eqref{HPF}, \eqref{EOM1} and \eqref{EOM2}. The result is
\begin{align}\label{coincidence0}
\left[\sig\right]=\left[\sigma_{;\mu}\right] = \left[\sigma_{;\mu^\prime}\right] = 0,
\end{align}
where the square brackets $\left[\ldots\right]$ indicate that the coincidence limit $x \to x^\prime$ has been taken. Finding coincidence limits for the second order covariant derivatives is a bit more involved but one finds (see \cite{SyngeGR})
\begin{gather}\label{coincidence}
\left[\sig_{\mu\nu}\right] = \left[\sig_{\mu^\prime \nu^\prime}\right] = g_{\mu\nu} \qquad \left[\sig_{\mu \nu^\prime}\right] = -g_{\mu\nu}.
\end{gather}
The covariant derivatives of $\sig(x,x')$, being the derivatives of a bi--scalar, transform as contravariant vectors. In particular, $\sig^\mu(x,x')$ transforms as a scalar at $x'$ and a contravariant vector at $x$, and vice versa for $\sig^{\mu'}(x,x')$. It follows that $\sig^{\mu'}(x,x')$ behaves exactly like the ``generalized coordinate'' mentioned in the preamble to this section and therefore can be used to define a covariant version of the Fourier kernel.

\subsection{Covariant Fourier Transform}
Fix a point $x' \in \mc{M}$ and let $p_{\mu'} \in T_{x'}\mc{M}$. The combination $p_{\mu'} \sig^{\mu'}(x,x')$ transforms like a scalar at both $x$ and $x'$ and so a natural candidate for the covariant Fourier kernel is given by $\exp(ip_{\mu'}\sig^{\mu'}(x,x'))$. This form of the kernel is not functional though since it leaves implicit the dependence of $p_{\mu'}$ on $x'$. Instead, we make use of the tetrad $e^a_{\mu'}$ and write $p_{\mu'} = p_a e^a_{\mu'}(x')$, so that the desired kernel is given by
\begin{align}\label{FK}
\exp \left[ip_a e^a_{\mu'}(x') \sig^{\mu'}(x,x')\right].
\end{align}
\indent The covariant Fourier transform (see \cite{covariantfourier, covariantfourier2} for an earlier implementation) is now defined as a map, $\mc{F}_{x'}$, from scalar functions on $\mathcal{M}$ to scalar functions on the cotangent plane $T^*_{x'}\mathcal{M}$, i.e. 
\begin{align*}
\mathcal{F}_{x'}: f(x) &\mapsto \hat{f}_{x'}(p),
\end{align*}
where
\begin{align}\label{FT}
\hat{f}_{x'}(p) \equiv \int d\mu(x) \mathcal{V}^{1/2}(x,x')\exp\left(-ip_ae^a_{\mu'}(x')\sig^{\mu'}(x,x')\right)f(x).
\end{align}
The factor $\mathcal{V}(x,x')$ is a bi--scalar known as the Van--Vleck Morette determinant (\cite{VanVleck},\cite{Morette}), and arises as the Jacobian in the change of variables $x^\mu \mapsto \sig^{\mu'}(x,x')$. It is given explicitly by
\begin{align}
\mathcal{V}(x,x') = g^{-1/2}(x)\left\vert\det\left(\sig_{\mu\nu'}(x,x')\right)\right\vert g^{-1/2}(x'),
\end{align}
and satisfies $[\mc{V}] = 1$.
We can also define an inverse Fourier transform via\footnote{For sake of convenience we are not including factors of $(2\pi)^4$}
\begin{align}
\mathcal{F}_{x'}^{-1}(\hat{f}_{x'})(x) \equiv \int d\nu(p)\mathcal{V}^{1/2}(x,x') \exp\left(ip_ae^a_{\mu'}(x')\sig^{\mu'}(x,x')\right)\hat{f}_{x'}(p),
\end{align}
where $d\nu(p) = g^{-1/2}(x')d^4p$. A few technical remarks regarding the domain of the covariant Fourier transform are in order. The definition of the worldfunction assumes the existence of a \textit{unique} geodesic connecting $x$ to $x'$; a condition which is not, in general, satisfied for two arbitrary points in $\mc{M}$. To ensure that the worldfunction remains single valued its domain must be restricted to a normal convex neighbourhood of $x'$, denoted $C_{x'}$. Consequently, the Fourier transform must be restricted to functions which vanish outside of $C_{x'}$. We must similarly restrict the inverse Fourier transform to functions in the image of $\mc{F}_{x'}$. For a more detailed discussion we refer the reader to \cite{scalarQFT}.\\
\indent A straightforward calculation now gives the following representation for the delta function on $\mathcal{M}$   
\begin{align}\label{deltaM}
\delta(x,y) = \int d\nu(p) \mathcal{V}^{1/2}(x,x')\mathcal{V}^{1/2}(y,x') \exp\left[ip_a e^a_{\mu'}\left(\sig^{\mu'}(x,x') - \sig^{\mu'}(y,x')\right)\right],
\end{align} 
where we have assumed that $\delta(x,y)$ is ``covariant'' in that it satisfies
\begin{align*}
\int d\mu(x) \delta(x,y) f(x) = f(y),
\end{align*}
for some scalar function $f(x)$.  The delta function presented in \eqref{deltaM} emphasizes the symmetry between $x$ and $y$ but observing that $\mathcal{V}(x,x')$ and $\mathcal{V}(y,x')$ can be factored out of the integral allows us to write
\begin{align}\label{delta}
\delta(x,y) = \int d^4p \mathcal{V}(y,x') \exp\left[ip_a e^a_{\mu'}\left(\sig^{\mu'}(x,x') - \sig^{\mu'}(y,x')\right)\right].
\end{align}
This formulation is useful when considering the special case $y = x'$ for which the delta function simplifies to
\begin{align}\label{deltafixed}
\delta(x,x') = \int d^4p\exp\left[ip_a e^a_{\mu'}\sig^{\mu'}(x,x') \right].
\end{align}
Note that the delta function, $\delta(x,x')$, will always have a well defined covariant Fourier transform since it is guaranteed to vanish outside of $C_{x'}$.

\section{Interaction Vertex in Curved Spacetime}
Having concluded our development of the covariant Fourier transform we are now prepared to continue with  the programme suggested at the conclusion of Section \ref{sec:Worldline Curved}.
\subsection{Implementing Localization}

Recall our set-up: An arbitrary process is represented by an oriented graph $\Gamma$ with local interactions and relevant amplitude given by
\begin{align}
\mc{I}_\Gamma(x_e^\mathrm{in},x_{-e}^\mathrm{out}) &=\int\prod_{v \in \Gamma}d\mu(z_v) \prod_{\mathrm{initial}\; e} \mc{G}_e\left(x_e, z_{e,t}\right)\prod_{\mathrm{final}\; e}\mc{G}_e\left(z_{e,s}, x_{-e}\right) \prod_{\mathrm{internal}\; e}\mc{G}_e\left(z_{e,s}, z_{e,t}\right).
\end{align}
To make locality explicit we extract a delta function for each edge sourcing or terminating at a vertex, viz
\begin{align}\label{Icurved}
\mc {I}_\Gamma(x_e^\mathrm{in},x_{-e}^\mathrm{out}) &=\int\prod_{v \in \Gamma}d\mu(z_v) \prod_{v,e}d\mu(x_e)\delta(x_e,z_v) \prod_{e \in \Gamma} G_e\left(x_e, x_{-e}\right).
\end{align}
Define $\widetilde{\mc{I}}_\Gamma$ to be the quantity obtained from $\mc{I}_\Gamma$ by dropping the vertex integrals and fixing the $z_v$ to be \textit{distinct} points in spacetime. Taking $x' = z_v$ in equation \eqref{deltafixed} then allows us to Fourier transform the delta functions appearing \eqref{Icurved}. We find
\begin{align*}
\widetilde{\mc{I}}_\Gamma(x_e^\mathrm{in},x_{-e}^\mathrm{out},z_v)  &= \int \prod_{v,e}  d \mu(x_e) \delta\left(x_e,z_v\right)\prod_{e\in \Gamma} G_e\left(x_e, x_{-e}\right)\\
&= \int\prod_{v,e}  d\mu(x_e) d^4p_{e}\exp\left(-ip_{a,e}e^a_{\mu_v}(z_v)\sig^{\mu_v}(x_e,z_v)\right)\prod_{e\in \Gamma} G_e\left(x_e, x_{-e}\right)\\
&= \int\prod_{v,e}  d\mu(x_e) d^4p_{e}\prod_{e\in \Gamma} \mathcal{D}\mu_e \exp(-iS_\Gamma),
\end{align*}
where the action $S_\Gamma$ is given by
\begin{align}\label{Scurved}
S_\Gamma = -\sum_{e \in \Gamma}\int_0^1 d\tau \left[\dot{X}^\mu_e P_{a,e}e^a_\mu - N_e\left(P_{a,e} P_{b,e}\eta^{ab} +m_e^2\right)\right] + \sum_{v,e} p_{a,e}e^a_{\mu_v}\sigma^{\mu_v}(x_e,z_v).
\end{align}
\indent As in the case of flat spacetime we obtain the vertex factor, along with the kinematical equations of motion, by taking the variation of $S_\Gamma$. The equations describing the free evolution of a particle were derived earlier (see \eqref{curvedP1} - \eqref{curvedP3}) and shown to be consistent with the geodesic equation. As such, we can focus on variations at the vertices which we find to be\footnote{We have assumed that $\delta X_e(0)=0$ and $\delta X_e(1) = 0$ for incoming and outgoing edges respectively.}
\begin{align*}
\sum_{v,e}&\left[e^a_{\mu_v}(z_v)\sig^{\mu_v}(x_e,z_v)\delta p_{a,e} + p_{a,e}e^a_{\mu_v,\nu_v}(z_v)\sig^{\mu_v}(x_e,z_v)\delta z^{\nu_v}_v + p_{a,e}e^a_{\mu_v}(z_v)\sig^{\mu_v}_{\pa \nu}(x_e,z_v)\delta x_e^\nu \right.\\
&\qquad \left. + p_{a,e}e^a_{\mu_v}(z_v)(\partial_{\nu_v}\sig^{\mu_v}(x_e,z_v))\delta z^{\nu_v}_v - P_{a,e}(0)e^a_\nu(x_e)\delta x^\nu_e \right].
\end{align*}
Setting the variations to zero we obtain the relevant equations of motion
\begin{gather}
e^a_{\mu_v}(z_v)\sig^{\mu_v}(x_e,z_v) = 0 \qquad \forall v \in \Gamma \label{local}\\
P_{a,e}(0)e^a_{\mu}(x_e) = p_{a,e}\N_{x^\mu_e}\sig^a(x_e,z_v)\qquad \forall e \in \Gamma\label{Pp}\\
\sum_{\substack{{s_e = v}\\{t_e = v}}}p_{a,e}\N_{z_v^\mu}\sig^{a}(x_e,z_v)= 0 \qquad \forall v \in \Gamma\label{constraint},
\end{gather}
where we have made use of the notation $\sig^a(x,x') = \sig^{\mu'}(x,x')e^a_{\mu'}(x')$. Note also that whenever $x_e$ and $z_v$ appear in the same equation the edge $e$ is assumed to innervate the vertex $v$. From \eqref{sigmu} we see that the first of these equations requires $x_e = z_v$, which is just the locality condition. Taking the coincidence limit on either side of the remaining equations, making use of \eqref{coincidence0} and \eqref{coincidence} and multiplying by the inverse tetrad we find
\begin{align}
P_{a,e}(0) = -p_{a,e} \quad \mathrm{and} \quad \sum_{\substack{{s_e = v}\\{t_e = v}}}p_{a,e} = 0.
\end{align}
These equations should be supplemented with the equation for conservation of momenta along an edge. As shown in \eqref{Pedge} the momenta $P_e(1) = -P_{-e}(0)$ at the end of an edge is related to the initial momenta $P_e(0)$ by parallel transport along $e$, denoted $U_{e}\equiv \overrightarrow{\exp}(\int_{e} \rd x^{\mu} \omega_{\mu})$. Thus, the equation governing conservation of momentum along an edge is given by 
\be\label{local1}
p_{-e}^{a}+ (p_{e }\!\cdot\! U_{e})^{a}=0,
\ee
where we denote $(p\cdot U)^{b}= p^{a}U_{a}{}^{b}$. As in the case of flat spacetime we can use the localization condition to relate the vertex momenta to a difference in position, although here the computation is more subtle. Begin with the derivative of the worldfunction evaluated at the endpoints of the edge $e$, i.e. $\sig^{\mu_{x_e}}(x_{-e},x_e)$. Equation \eqref{sigmup} then allows us to write 
\begin{align*}
\sig^{\mu_{x_e}}(x_{-e},x_e) &= -(s_1 - s_0)\partial_s X_e^{\mu}(0)\\
&= -\partial_s X_e^{\mu}(0) \int_0^1 N_e(\tau) d\tau,
\end{align*}
where $ds = Nd\tau$ is the proper time along the world line. Now use \eqref{curvedP2} to replace $\partial_s X^{\mu}_e(0)$ in favour of $P_e(0)$ so that
\be
\sig^{\mu_{x_e}}(x_{-e},x_e) = -\tau_e P_{e}^{a}(0)e^{\mu}_a(x_e), 
\ee
where $\tau_e = \int N_ed\tau$. Finally, the localization equation allows us to identify $x_{-e} = z_{t_e}$ and $x_e = z_{s_e}$ while our relation between edge and vertex momentum yields $P_{b,e}(0) = -p_{a,e}$ and so
\be
\sigma^{a}(z_{t_{e}},z_{s_{e}})  =  \tau_{e}  p_{e}^{a}.
\ee
These localization equations are compatible with the parallel transport equation of momenta along an edge since
\be\label{reverse}
\sigma^{a}(z_{t_{e}},z_{s_{e}}) U_{e a}{}^{b} =-  \sigma^{b}(z_{s_{e}},z_{t_{e}});
\ee
see Appendix \ref{ap:propagator} for a proof.

\subsection{Localisation on Loops}
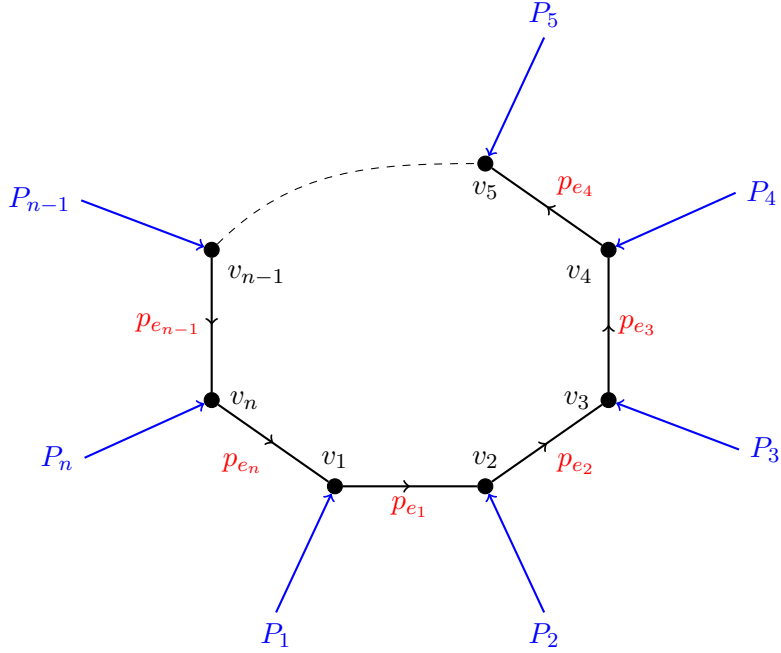
\begin{figure}[H]
\begin{center}
\begin{tikzpicture}
[place/.style={circle, draw= black, fill=black, inner sep =2pt}]
\node[place] (a) at (0,0) [label = above: $v_1$]{};
\node[place] (b) at (2,0)[label = above: $v_2$]   {};
\node[place] (c) at (3.638,1.147)[label = west: $v_3$]  {};
\node[place] (d) at (3.638,3.147)[label = south west: $v_4$]  {};
\node[place] (e) at (2,4.294)[label = below: $v_5$]  {};
\node[place] (f) at (-1.638,3.147)[label = south east: $v_{n-1}$]  {};
\node[place] (g) at (-1.638,1.147)[label = east: $v_n$]  {};
\draw[->-, thick] (a) to node {} node [auto, swap, red] {$p_{e_1}$} (b);
\draw[->-, thick] (b) to node {} node [auto, swap, red] {$p_{e_2}$} (c);
\draw[->-, thick] (c) to node {} node [auto, swap, red] {$p_{e_3}$} (d);
\draw[->-, thick] (d) to node {} node [auto, swap, red] {$p_{e_4}$} (e);
\draw[->-, thick] (f) to node {} node [auto, swap, red] {$p_{e_{n-1}}$} (g);
\draw[->-, thick] (g) to node {} node [auto, swap, red] {$p_{e_{n}}$} (a);
\draw[-,dashed] (f) to [out=45,in=180] (e);
\node (aa) at (-0.845,-1.813) {};
\node (bb) at (2.845,-1.813) {};
\node (cc) at (5.508,0.438) {};
\node (dd) at (5.463,3.966) {};
\node (ee) at (2.845,6.107) {};
\node (ff) at (-3.508,3.856) {};
\node (gg) at (-3.463,0.318) {};
\draw[->, thick, blue] (aa) to node  [blue, at start, below]{$P_1$}   (a);
\draw[->, thick, blue] (bb) to node  [blue, at start, below]{$P_2$} (b);
\draw[->, thick, blue] (cc) to node  [blue, at start, right]{$P_3$}(c);
\draw[->, thick, blue] (dd) to node  [blue, at start, right]{$P_4$}(d);
\draw[->, thick, blue] (ee) to node  [blue, at start, above]{$P_5$}(e);
\draw[->, thick, blue] (ff) to node  [blue, at start, left]{$P_{n-1}$}(f);
\draw[->, thick, blue] (gg) to node  [blue, at start, left]{$P_n$}(g);
\end{tikzpicture}
\caption{\textit{The loop $L=e_1e_2\cdots e_n$ }}
\label{fig:loop}
\end{center}
\end{figure}

In this section we will study the localization equations (\ref{local1}) for a graph that possesses a loop $L$. Assume that  $L$ consists of the edges $L=e_{1}e_{2}\cdots e_{n}$, and that $e_{i}=(ii+1)$, goes from vertex $i$ to vertex $i+1$. 
We denote by $P_{i}$ the external momenta incoming to vertex $i$ and by $ p_{e_{i}}=p_{ii+1}$ the momenta on edge $e_{i}$ starting at vertex $i$. This is illustrated in Figure \ref{fig:loop}. The localization equations, \eqref{local1}, read
\be\label{local loop}
p_{i+1i} + p_{ii+1}\!\cdot\! U_{ii+1} =0,\qquad  P_{i}= p_{ii-1}+ p_{ii+1},
\ee
where $i = 1,\ldots, n$ and addition is modulo $n$. Define $U_{aa+m}\equiv U_{aa+1}U_{a+1a+2}\cdots U_{a+m-1 a+m}$ to be the holonomy from $a$ to $a+m$, so that upon summing the above relation we obtain
\be\label{S}
S_{n}\equiv P_{n}+ P_{n-1}\!\cdot\!U_{n-1n}+\cdots P_{1}\!\cdot\!U_{1n} = p_{n1}\!\cdot\!(1-H_{n}),
\ee
where $H_{n}= U_{n1}U_{12}\cdots U_{n-1n}$ is the total holonomy around the loop based at the vertex $n$. To generalize this relation to an arbitrary base vertex we introduce the momenta transported from the vertex $i$ 
\be
\hat{P}_{i}\equiv  P_{i}\!\cdot\!U_{in},
\ee 
so that $S_{n}=\sum_{i=1}^{n}\hat{P}_{i}$. We immediately obtain $p_{ii+1} U_{ii+1} = \hat{P}_i + p_{i-1i}U_{i-1i+1}$ which can then be solved iteratively to express $p_{ii+1}$ in terms of the external momenta and $p_{n1}$ as
\be\label{pi}
p_{ii+1}\!\cdot\!U_{in} = (\hat{P}_{i} +\cdots \hat{P}_{1}) + p_{n1}H_{n}.
\ee
Putting equations \eqref{S} and \eqref{pi} together we see that the loop momenta $p_{ii+1}$ are entirely determined by the external momenta, which is related to the fact that in  presence of gravity the total momenta around a loop is no longer conserved.
When spacetime is flat $H_n = 1$ and so, the right hand side of \eqref{S} vanishes, total momentum is  conserved and the loop momentum is independent of the external momenta. Consequently, one must integrate over the loop momenta when performing the path integral, leading to the well known problems with ultraviolet divergences. On the other hand, when gravity is present the holonomy will differ from the identity allowing, quite generically, the operator $(1-H_{n})$ to be inverted. In this case we can express all the momenta in terms of the external ones! For a small loop the holonomy approximates to $H^{a}{}_{b}= \delta^{a}_{b} +  R^{a}{}_{b\mu\nu}\Delta A^{\mu\nu}+\cdots$ where $\Delta A$ is the loop area. The invertibility of $(1-H)$ is therefore related to the invertibility 
of $ R^{a}{}_{b}(X)$ for all invertible bivectors $X$. Thus, in a fully curved background the only way (generically) to have a non invertible $(1-H)$ is to consider a loop of zero extension, i.e. with $\Delta A=0$. It is these loops of zero size that give rise to divergences in quantum field theory.\\
\indent In summary, the effect of a gravitational field on an extended loop is to produce a violation of total momentum conservation. This phenomena is related to the fact that, in the present case, loop momenta can be expressed entirely in--terms of the external momenta. Therefore, if we could argue that quantum gravity requires the expectation value of the holonomy $\langle H\rangle$ to be different from unity for all loops, even those which shrink to an effective size of zero,
this would have a dramatic regulating effect on Feynman integrals, at least on their semi--classical evaluation. In particular, we could restrict loop integrals to a finite region of momentum space.
%

\subsection{Nonlocal vertex}
In the previous section we showed that coupling to a classical gravitational field modifies the loop propagator by introducing a holonomy (around the loop) into the conservation of momentum equation. On the other hand, we saw that momentum conservation at the vertices is unaffected, being identical to the relations derived for flat spacetime.\\
\indent A pertinent question arises, how do quantum gravity effects alter particle physics amplitudes? It is well known that the inclusion of quantum gravity introduces a new mass scale into the theory, namely the Plank mass. Our question can then be phrased more formally as follows: Suppose we couple gravity to a Feynman integral and compute, by some method, the quantum gravity average, how does this evaluation affect the Feynman integral? It is tempting to assume that the computation,  done in any theory of quantum gravity, will correspond to a mass dependent deformation of the standard integral. According to the philosophy presented here, and assuming that new degrees of freedom do not appear, this deformation can in turn be entirely reabsorbed into a deformation of the particle action. \\
\indent It is natural to assume that this deformation will affect the vertex interaction. Indeed it was the vertex factor $p_{a}\sigma^{a}(x_{e},z_{v})$ which, as we have seen, determined the localization condition $x_{e}=z_{v}$. Such exact localization will certainly be relaxed in a theory of quantum gravity. We  propose, therefore, to modify the vertex interaction as an effective way to include quantum gravity (de--localizing type) effects. \\
\indent The simplest such modification is to consider a vertex interaction of the form
$ p_{a}\sigma^{a}(x_{e},z_{v}) - p_{a}p^{a}/2M$ where $M$ is the quantum gravity mass scale.
In the Euclidean formulation of the theory this amounts to replacing the vertex interaction, $\delta(x,z)$, by a Gaussian weight 
\be
\delta_{M}(x,z)= \left(\frac{M}{2\pi}\right)^{\frac{d}2} e^{-M\sigma(x,z)}.
\ee
The equations of motion resulting from this de--localized vertex are readily found to be (c.f. \eqref{local} - \eqref{constraint})
\bea
M \sig^{a}(x_e,z_v) &=& p^{a}_{e}  \label{local2}\\
P_{a,e}(0)e^a_{\mu}(x_e) &=& p_{a,e}\N_{x_{e}^{\mu}}\sig^{a}_{\pa }(x_e,z_v)
\label{Pp2}\\
\sum_{\substack{{s_e = v}\\{t_e = v}}}\left[  p_{a,e} \N_{z_{v}^{\mu}}\sig^{a}_{\pa}(x_e,z_v)\right]&=& 0 
\label{constraint2},
\eea
Observe that the mass scale only enters in the first equation, modifying the locality condition by ensuring that $x_{e}$ and $z_{v}$ are no longer identified. Substituting this into equation \eqref{constraint2} and using that the Synge function satisfies\footnote{See equation \eqref{SD2}} $\sig^{a}\N_{z^\mu}\sig_{a}(x,z)= e_{\mu}^{a}(z) \sig_{a}(x,z)$ we obtain
\be
\sum_{\substack{{s_e = v}\\{t_e = v}}}p^{a}_{e}=0,
\ee
which, as before, is the usual conservation of vertex momenta. Where the modification becomes apparent is in the relationship between the endpoint momenta $P$ and the vertex momenta $p$; from equation \eqref{Pp2} we have
\be
P_{e}^{a}(0) = M \eta^{ab} e^\mu_{b}(x_e)\N_{x_{e}^{\mu}}\sig_{b}(x_e,z_v)
= M\eta^{ab}\sig_{\mu_e}(x_e,z_v)e^{\mu_e}_a(x_e).
\ee
Let $V_{ev}$ denote the parallel propagator from $z_{v}$ to $x_{e}$, then $\sig^{\mu_e} = - [V_{ev}]^{\mu_e}_{\pa \nu_v}\sig^{\nu_v}(x_e,z_v)$ and so
\begin{align}
P_{e}^{a}(0) &= -Me_{\mu_e}^a(x_e)[V_{ev}]^{\mu_e}_{\pa \nu_v}\sig^{\nu_v}(x_e,z_v)\\
&=-M[V_{ev}]^{a}_{\pa b}\sig^{b}(x_e,z_v)\\
&=-(p_e\!\cdot\! V_{ev})^a,
\end{align}
where we have made use of the notation $[V_{ev}]^a_{\pa b} = e^a_{\mu_e}(x_{e})[V_{ev}]^{\mu_e}_{\pa \nu_z}e^{\nu_z}_b(z_{v})$ in the second line. Remarkably, this implies that the conservation of momenta along edges is modified in a trivial manner, viz
\begin{align}\label{pdelocal}
p_{-e} + p_e \!\cdot\!(V_{s_e e}U_e V_{-et_e}) = 0.
\end{align}
The term in brackets is the full propagator from $s_e$ to $t_e$ indicating that the form of this equation is identical to the one consider for a local vertex, see \eqref{local1}. It follows that the de--localization of the vertex does not affect the momenta conservation equations, either at the vertex or along the edges. It's only effect is to modify the relationship between momenta and coordinates, e.g. in a flat spacetime the modification enters through a shift in proper time
\begin{align}
z_{t_e} - z_{s_e} = (\tau_e +1/2M)p_e.
\end{align}
A more general modification of the vertex, that is quadratic, Lorentz invariant and symmetric under exchanges of momenta will include an additional term proportional to $\sum_{e,e'}p^a_{e'}p_{e,a}/2M$. 
This term gives rise to effects which are similar to the ones considered above.
\section{Conclusion}
In this paper we considered the extent to which demanding local interactions affects the form of the interaction vertex. We began with flat spacetime, showing that localization gives rise to conservation of momentum at each vertex and the identification of edge and vertex momentum. Generalizing to a curved geometry we found that conservation of vertex momentum was maintained but now edge and vertex momentum were no longer identical, instead being related by parallel transport. Next, we considered the effects on loop diagrams immersed in a gravitational field, showing that the loop momenta is entirely determined by momenta external to the loop. This result presents an intriguing approach for regularizing loop integrals in standard QFT. Finally, we argued that the semi--classical effects of quantum gravity could be accounted for by modifying the interaction vertex so as to relax strict locality. Making a particular choice of de--localized vertex we demonstrated, rather remarkably, that conservation of momentum at a vertex was preserved; in fact the only modification was in the relationship between edge and vertex momenta.
This opens the way towards a deeper investigation of how to de--localize the vertex in the presence of quantum gravity and whether this de--localization is related to introducing non--trivial momentum space geometry as postulated in Relative locality.

\section*{Acknowledgments}
We would like to thank J. Kowalski-Glikman for sharing his results during his visits at PI and L. Smolin for encouragements.
This research was supported in part by Perimeter Institute for Theoretical Physics. Research at Perimeter Institute is supported by the Government of Canada through Industry Canada and by the Province of Ontario through the Ministry of Research and Innovation. This research was also partly supported by grants from NSERC.

\appendix
\section{Parallel Propagator}\label{ap:propagator}
Consider two points in spacetime $x,x' \in \mc{M}$ joined by a geodesic $\gamma_{xx'}$.  The parallel propagator, denoted $U^\mu_{\pa \mu'}(x,x')$, is the operator which takes a vector field at $x'$ and parallel transports along $\gamma_{xx'}$ to a vector field at $x$. By definition, a geodesic is a curve which parallel transports its own tangent vector, i.e. $T^\mu \N_\mu T_\nu = 0$, where $T^\mu = d\gamma_{xx'}^\mu/d\tau$ for some affine parameter $\tau$. In terms of the parallel propagator this becomes
\begin{align}\label{tangents}
T^\mu(x) = U^\mu_{\pa \mu'}(x,x')T^{\mu'}(x), \qquad T^{\mu'}(x') = U^{\mu'}_{\pa \mu}(x,x') T^\mu(x).
\end{align}
The tangent vectors $T^\mu(x)$ and $T^{\mu'}(x')$ are related to the derivative of the worldfunction at $x$ and $x'$ via equations \eqref{sigmu} and \eqref{sigmup}, respectively. In particular,
\begin{align}
\sig^\mu(x,x') = T^\mu = U^\mu_{\pa \mu'}T^{\mu'}(x') = -U^\mu_{\pa \mu'} \sig^{\mu'}(x,x'),
\end{align}
with a similar computation holding for $\sig^{\mu'}$. Substituting into \eqref{tangents} we find
\begin{align}
\sig^\mu(x,x') + U^{\mu}_{\pa \mu'}(x,x') \sig^{\mu'}(x,x') &= 0\label{Usig1}\\
\sig^{\mu'}(x,x') + U^{\mu'}_{\pa \mu} (x,x') \sig^{\mu}(x,x') &= 0 \label{Usig2},
\end{align} 
which justifies the statement made in the text that, when acting on $\sig_\mu$ or $\sig_{\mu'}$, the second order mixed derivative of the worldfunction behaves like the parallel propagator (up to a sign). These equations can be written in the $e^a$ basis by making a couple of observations. Focusing on \eqref{Usig1}, we have 
\begin{gather}
\sig^\mu(x,x') = \N_{x_\mu} \sig(x,x') = \N_{x_\mu}\sig(x',x) \quad \Longrightarrow \quad  e_\mu^a(x)\sig^\mu(x,x') = \sig^a(x',x),\\
U^{\mu}_{\pa \mu'} \sig^{\mu'} = U^{\mu}_{\pa \mu'}e^{\mu'}_a(x') e^a_{\nu'}(x')\sig^{\nu'} = U^{\mu}_{\pa \mu'}e^{\mu'}_a(x') \sig^{a},
\end{gather}
with similar results holding for \eqref{Usig2}. Thus, we obtain the form of these equations used in the text
\begin{align}
\sig^a(x',x) + U^{a}_{\pa b}(x,x') \sig^{b}(x,x') &= 0,\label{Usig1e}\\
\sig^{a}(x,x') + U^{a}_{\pa b} (x,x') \sig^{b}(x',x) &= 0 \label{Usig2e}.
\end{align}

\bibliography{Final_Worldline}{}

\end{document}